\documentclass[entropy,article,submit,moreauthors]{Definitions/mdpi} 

\firstpage{1} 
\pubvolume{1}
\issuenum{1}
\articlenumber{0}
\pubyear{2020}
\copyrightyear{2020}
\history{Received: date; Accepted: date; Published: date}

\newcommand{\sign}{\mathrm{sign}}
\newcommand{\vw}{\mathbf{w}}
\newcommand{\vW}{\mathbf{W}}

\newcommand{\vb}{\mathbf{b}}
\newcommand{\vr}{\mathbf{r}}
\newcommand{\vx}{\mathbf{x}}
\newcommand{\vxi}{\boldsymbol{\xi}}
\newcommand{\vo}{\mathbf{o}}
\newcommand{\extr}{\mathop{extr}}

\Title{Aspects of a phase transition in high-dimensional random geometry}

\Author{Axel Pr\"user $^{1}$*, Imre Kondor $^{2,3,4}$ and Andreas Engel $^{1}$}

\AuthorNames{Axel Pr\"user, Imre Kondor and Andreas Engel}

\address{
$^{1}$ \quad Institute of Physics, Carl von Ossietzky University of Oldenburg, D-26111 Oldenburg, Germany\\
$^{2}$ \quad Parmenides Foundation, Pullach, Germany\\
$^{3}$ \quad London Mathematical Laboratory, London, UK\\
$^{4}$ \quad Complexity Science Hub, Vienna, Austria}

\corres{Correspondence: axel.prueser@uol.de}

\abstract{A phase transition in high-dimensional random geometry is analyzed as it arises in a variety of problems. A prominent example is the feasibility of a minimax problem that represents the extremal case of a class of financial risk measures, among them the current regulatory market risk measure Expected Shortfall. Others include portfolio optimization with a ban on short selling, the storage capacity of the perceptron, the solvability of a set of linear equations with random coefficients, and competition for resources in an ecological system. These examples shed light on various aspects of the underlying geometric phase transition, create links between problems belonging to seemingly distant fields and offer the possibility for further ramifications.}



\begin{document}


\section{Introduction}
A large class of problems in random geometry is concerned with the collocation of points in high-dimensional space. Applications range from optimization of financial portfolios \cite{Kondor2007Noise}, binary classifications of data strings \cite{Cover} and optimal stategies in game theory \cite{BeEn} to the existence of non-negative solutions to systems of linear equations \cite{LaEn,Garnier-Brun}, the emergence of cooperation in competitive ecosystems \cite{MacArthur,TiMo}, and linear programming with random parameters \cite{MinimaxTodd}. It is frequently relevant to consider the case where both the number of points $T$ and the dimension of space  $N$ tend to infinity. This limit is often characterized by abrupt qualitative changes reminiscent of phase transitions when an external parameter or the ratio $T/N$ vary and cross a critical value. At the same time, this high-dimensional case is amenable to methods from the statistical mechanics of disordered systems offering additional insight. 

Some results obtained in different disciplines are closely related to each other without the connection always being appreciated. In the present paper we discuss some particular cases. We will show that the boundedness of the expected maximal loss as well as the possibility of zero variance of a random financial  portfolio is closely related to the existence of a linear separable binary coloring of random points called a dichotomy. Moreover we point out the connection with the existence of non-negative solutions to  systems of linear equations and with mixed stategies in zero-sum games. On a more technical level and for the above mentioned limit of large instances in high-dimensional spaces we also make contact between replica calculations performed for different problems in different fields. 

In addition to uncovering the common random geometrical background of seemingly very different problems, our comparative analysis sheds light on each of them from various angles and points to ramifications in their respective fields.



\section{Dichotomies of random points}
Consider an $N$-dimensional Euclidean space with a fixed coordinate system. Choose $T$ points in this space and color them either black or white. The coloring is called a dichotomy if a hyperplane through the origin of the coordinate system exists that separates black points from white ones.

To avoid special arrangements like all points falling on one line the points are required to be in what is called general position: the position vectors of any subset of $N$ points should be linearly independent. Under this rather mild prerequisite the number $C(T,N)$ of dichotomies of $T$ points in $N$ dimensions only depends on $T$ and $N$ and not on the particular location of the points. This remarkable result was proven in several works among them a classical paper by Cover \cite{Cover}. Establishing a recursion relation for $C(T,N)$ he derived the explicit result

\begin{equation}
 C(T,N)=2\sum_{i=0}^{N-1}\binom{T-1}{i}.
\end{equation} 

\begin{figure}[H]
\centering
\includegraphics[width=8cm]{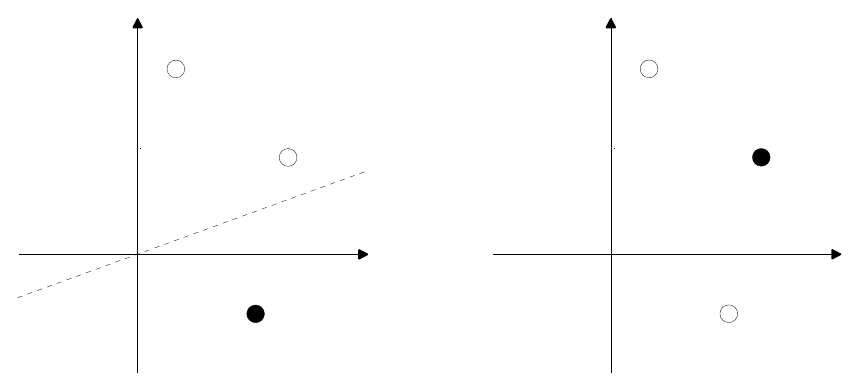}
\caption{Two colorings of three points in two dimensions. In the left one black and white points can be separated by a line through the origin; this coloring therefore represents a dichotomy. For the right one no such separating line exists.\label{fig:dich}}
\end{figure}   

\begin{figure}[H]
\centering
\includegraphics[width=7cm]{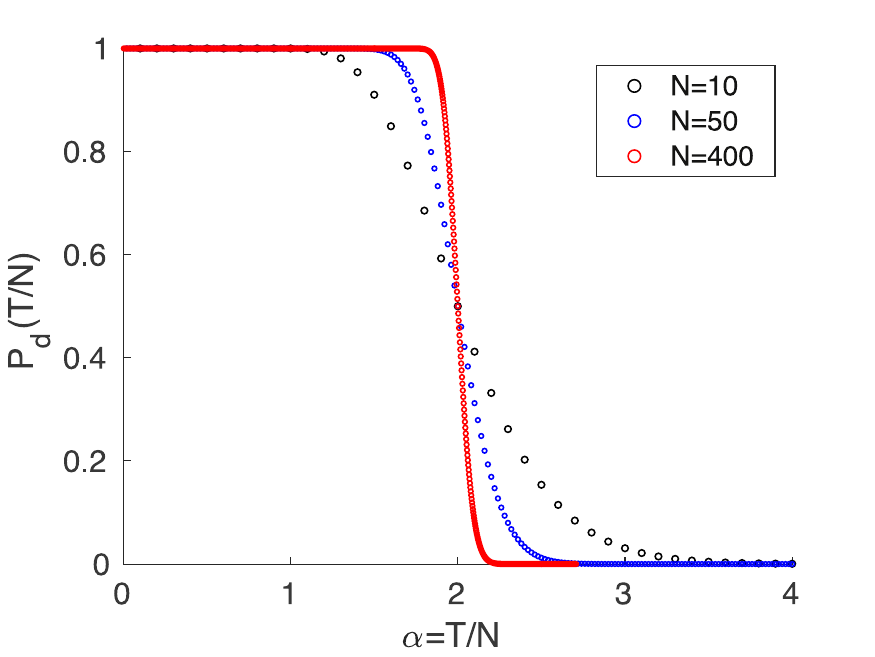}
\caption{Probability $P_\mathrm{d}(T,N)$ that $T$ randomly colored points in general position in $N$-dimensional space form a dichotomy as function of the ratio $\alpha$ between $T$ and $N$ for different values of $N$. The transition between the limiting values $P=1$ at $\alpha=1$ and $P=0$ at large $\alpha$ becomes increasingly sharp when $N$ grows. \label{fig:cover}}
\end{figure}

If the coordinates of the points are chosen at random from a continuous distribution the points are in general position with probability one. Since there are in total $2^T$ different binary colorings of these points and only $C(T,N)$ of them are dichotomies we find for the probability that $T$ random points in $N$ dimensions with random coloring form a dichotomy the cumulative binomial distribution 

\begin{equation}\label{rescov}
 P_\mathrm{d}(T,N)=\frac{C(T,N)}{2^T}=\frac{1}{2^{T-1}}\sum_{i=0}^{N-1}\binom{T-1}{i}.
\end{equation} 
Hence $P_\mathrm{d}(T,N)=1$ for $T\leq N$, $P_\mathrm{d}(T,N)=1/2$ for $T=2N$ and $P_\mathrm{d}(T,N)\to 0$ for $T\to\infty$. The transition from $P\simeq 1$ at $T=N$ to $P\simeq 0$ at large $T$ becomes sharper with increasing $N$. This is clearly seen when considering the case of constant ratio

\begin{equation}\label{defal}
 \alpha:=\frac{T}{N}
\end{equation} 
between the number of points and the dimension of space for different values of $N$, which shows an abrupt transition at $\alpha_c=2$ for $N \to 
\infty$, cf. Fig.~\ref{fig:cover}.

For later convenience it is useful to reformulate the condition for a certain coloring to be a dichotomy in different ways. Let us denote the position vector of point $t, t=1,...,T$, by $\vxi^t\in\mathbb{R}^N$ and its 
coloring by the binary variable $\zeta^t=\pm 1$. If a separating hyperplane exists it has a normal vector $\vw\in\mathbb{R}^N$ that fulfills

\begin{equation}\label{defdich}
 \zeta^t=\sign(\vw\cdot\vxi^t),\qquad t=1,...,T,
\end{equation} 
where we define $\sign(x)=1$ for $x\geq 0$ and $\sign(x)=-1$ otherwise. With the abbreviation

\begin{equation}\label{defr}
 \vr^t:=\zeta^t\vxi^t
\end{equation} 
eq.~\eqref{defdich} translates into $\vw\cdot\vr^t\geq 0$ for all $t=1,...,T$ which for points in general position is equivalent to the somewhat stronger condition

\begin{equation}\label{version1}
 \vw\cdot \vr^t>0, \qquad t=1,...,T.
\end{equation} 
A certain coloring $\zeta^t$ of points $\vxi^t$ is hence a dichotomy if a vector $\vw$ exists such that \eqref{version1} is fulfilled, i.e. if its scalar product with all vectors $\vr^t$ is positive. This is quite intuitive since by going from the vectors $\vxi^t$ to $\vr^t$ according to the \eqref{defr} we replace all points colored black by their white colored mirror images (or vice versa). If we started out with a dichotomy, after the transformation all points will lie on the same side of the separating hyperplane. The meaning of Eq.~\eqref{version1} is clear: For $T$ random points in $N$ dimensions with coordinates chosen independently from a symmetric distribution there exists with probability $P_d(T,N)$ a hyperplane such that all these points lie on the same side of the hyperplane. This formulation will be crucial in section \ref{portfolios} to relate dichotomies to bounded cones characterizing financial portfolios.

Singling out one particular point $s=1,...,T$ this in turn implies that there is for any choice of $s$ a vector $\vw$ with 

\begin{equation}\label{h5}
 \vw\cdot \vr^t>0, \quad t=1,...,T, t\neq s\qquad\mathrm{and}\qquad \vw \cdot(-\vr^s)<0.
\end{equation} 
Consider now all vectors $\bar{\vr}$ of the form 

\begin{equation}\label{nncone}
 \bar{\vr}=\sum_{t\neq s} c^t \vr^t, \qquad\mathrm{with}\qquad c^t\geq 0,\;  t=1,...,T, t\neq s,
\end{equation} 
i.e., all vectors that may be written as linear combination of the $\vr^t$ with $t\neq s$ and all expansion parameters $c^t$ being non-negative. The set of these vectors $\bar{\vr}$ is called the {\em non-negative cone} 
of the $\vr^t, t\neq s$. Eq.~\eqref{h5} then means that $-\vr^s$ cannot be an element of this non-negative cone. This is clear since the hyperplane perpendicular to $\vw$ separates $-\vr^s$ from this very cone, an observation that is known as Farkas' lemma \cite{farkas}. Therefore, if a set of vectors $\vr^t$ forms a dichotomy no mirror image $-\vr^s$ of any of them 
may be written as a linear combination of the remaining ones with non-negative expansion coefficients

\begin{equation}\label{version2}
 \sum_{t\neq s} c^t \vr^t\neq -\vr^s, \qquad\forall c^t\geq 0.
\end{equation} 
Finally, adding $\vr^s$ to both sides of \eqref{version2} we find

\begin{equation}\label{version3}
 \sum_t c^t \vr^t\neq\vo, \qquad\mathrm{with}\qquad c^t\geq 0,\, t=1,...,T,\quad\mathrm{and}\quad
      \sum_t c^t>0,
\end{equation} 
where $\vo$ denotes the null vector in $N$ dimensions. Given $T$ points $\vr^t$ in $N$ dimensions forming a dichotomy it is therefore impossible to find a nontrivial linear combination of these vectors with non-negative 
coefficients that equals the null vector. 

Also this corollary to the Cover result is easily understood intuitively. 
Assume there were some coefficients $c^t\geq 0$ that are not all at the same time zero and that realize

\begin{equation}\label{h3}
 \sum_t c^t \vr^t=\vo.
\end{equation} 
If the points $\vr^t$ form a dichotomy then according to \eqref{version1} there is a vector $\vw$ that makes a positive scalar product with all of them. Multiplying \eqref{h3} with this vector we immediately arrive at a contradiction since the l.h.s. of this equation is positive and the r.h.s. is zero. 

Note that also the inverse of \eqref{version3} is true: if the points do not form a dichotomy a decomposition of the null vector of the type \eqref{h3} can always be found. This is related to the fact that the non-negative cone of the corresponding position vectors is the complete $\mathbb{R}^N$. For if there were a vector $\vb\in \mathbb{R}^N$ that lies not in this cone by Farkas' lemma there would be a hyperplane separating the 
cone from $\vb$. But the very existence of this hyperplane would qualify the points $\vr^t$ to be a dichotomy in contradiction to what was assumed.

In the limit $N\to \infty,\, T\to \infty$ with $\alpha=T/N$ staying constant the problem of random dichotomies can be investigated within statistical mechanics. To make this connection explicit, we first note that no inequality in \eqref{version1} is altered if $\vw$ is multiplied by a positive constant. To decide whether an appropriate vector $\vw$ fulfilling \eqref{version1} may be found or not it is hence sufficient to study vectors of a given length. It is convenient to choose this length as $\sqrt{N}$ requiring

\begin{equation}\label{sphernorm}
 \sum_{i=1}^N w_i^2=N.
\end{equation} 
Next, we introduce for each realization of the random vectors $\vr^t$ an energy function 

\begin{equation}
 E(\vw):=\sum_{t=1}^T \Theta\left(-\sum_i w_i r_i^t\right),
\end{equation} 
where $\Theta(x)=1$ if $x>0$ and $\Theta(x)=0$ otherwise is the Heaviside step function. This energy is nothing but the number of points violating \eqref{version1} for given vector $\vw$. Our central quantity of interest is the entropy of the groundstate of the system, i.e., the logarithm of the fraction of points on the sphere defined by \eqref{sphernorm} that realize zero energy

\begin{align}\label{defGaEn}
  S(\kappa, \alpha):=\lim_{N\to\infty}\frac{1}{N}\ln\frac{ \int \prod_{i=1}^{N} dw_i \, \delta (\sum_{i}w_i^2-N ) \, \prod_{t=1}^{\alpha N} \Theta\left( \sum_{i} w_i r_i^t-\kappa \right)}
  { \int \prod_{i=1}^{N} dw_i \, \delta ( \sum_{i}w_i^2-N )}.
\end{align}
Here $\delta(x)$ denotes the Dirac $\delta$-function and we 
have introduced the positive stability parameter $\kappa$ to additionally 
sharpen the inequalities \eqref{version1}. 

The main problem in the explicit determination of $S(\kappa,\alpha)$ is its dependence on the many random parameters $r_i^t$. Luckily, for large values of $N$ deviations of $S$ from its typical value $S_\mathrm{typ}$ become extremely rare and, moreover, this typical value is given by the average over the realizations of the $r_i^t$

\begin{equation}
 S_\mathrm{typ}(\kappa,\alpha)=\langle\langle S(\kappa,\alpha)\rangle\rangle.
\end{equation} 
The calculation of this average was performed in a by now classical calculation \cite{Gardner} which gave rise to the result 

\begin{equation}\label{resGaEn}
 S_\mathrm{typ}(\kappa,\alpha)=\extr_q\left[\frac{1}{2}\ln(1-q)+\frac{q}{2(1-q)}
  +\alpha\int Dt \ln H\left(\frac{\kappa-\sqrt{q}t}{\sqrt{1-q}}\right)\right],
\end{equation} 
where the extremum is over the auxiliary quantity $q$ and we have used the shorthand notations

\begin{equation}\label{shorthand}
 Dt:=\frac{dt}{\sqrt{2\pi}} e^{-\frac{t^2}{2}}\qquad\mathrm{and}\qquad H(x):=\int_x^\infty Dt .
\end{equation} 
More details of the calculation may be found in the original reference and in chapter 6 of \cite{EnvB}. Appendix A contains some intermediate steps for a closely related analysis.

Studying the limit $q\to 1$ of \eqref{resGaEn} reveals 

\begin{equation}
 S_\mathrm{typ}(\kappa,\alpha)\begin{cases}
                              >-\infty & \mathrm{if}\qquad \alpha<\alpha_c(\kappa)\\
                              \to-\infty & \mathrm{if}\qquad \alpha>\alpha_c(\kappa),
                               \end{cases}
\end{equation} 
corresponding to a sharp transition from solvability to non-solvability at a critical value $\alpha_c(\kappa)$. For $\kappa=0$ on finds $\alpha_c=2$ in agreement with \eqref{rescov}, cf. Fig.~\ref{fig:cover}.

Note that Cover's result \eqref{rescov} holds for all values of $T$ and $N$ whereas the statistical mechanics analysis is restricted to the thermodynamic limit $N\to\infty$. On the other hand, the latter can deal with all values of the stability parameter $\kappa$ whereas no generalization of Cover's approach to the case $\kappa\neq 0$ is known. 


\section{Phase transitions in portfolio optimization under the variance and the maximal loss risk measure}
\label{portfolios}

\subsection{Risk measures}

The purpose of this subsection is to indicate the finance context, in which the geometric problem discussed in this paper appears. A portfolio is the weighted sum of financial assets. The weights represent the parts of the total wealth invested in the various assets. Some of the weights are allowed to be negative (short positions), but the weights sum to 1; this is called the budget constraint. Investment carries risk and higher returns usually carry higher risk. Portfolio optimization seeks 
a trade-off between risk and return by the appropriate choice of the portfolio weights. Markowitz was the first to formulate the portfolio choice as a risk-reward problem \cite{Markowitz1952Portfolio}. Reward is normally regarded as the expected return on the portfolio. Assuming return fluctuations to be Gaussian-distributed random variables, portfolio variance offered itself as the natural risk measure. This setup made the optimization of portfolios a quadratic programming problem, which, especially in the case of large institutional portfolios, posed a serious numerical difficulty in its time. Another critical point concerning variance as a risk measure was that variance is symmetric in gains and losses, whereas investors are believed not to be afraid of big gains, only big losses. This consideration led to the introduction of downside risk measures, starting already with the semivariance \cite{Markowitz1959}. Later it was recognized 
that the Gaussian assumption was not realistic, and alternative risk measures were sought to grasp the risk of rare but large events, and also to allow risk to be aggregated across the ever inscreasing and increasingly heterogeneous institutional portfolios. Around the end of the eighties Value at Risk (VaR) was introduced by JP Morgan \cite{Morgan1995Riskmetrics} and subsequently it was widely spread over the industry by their RiskMetrics methodology \cite{Riskmetrics1996}. VaR is a high quantile, a downside risk measure{\footnote{Note that in the literature the profit and loss axis is often reflected, so that losses are assigned positive sign. It is under this convention that VaR is a high quantile, rather than a low one.}}. It soon came under academic criticism for its insensitivity to the details of the distribution beyond the quantile, and for its lack of sub-additivity. Expected Shortfall (ES), the average loss above the VaR quantile appeared around the turn of the century \cite{Acerbi2001ESasatool}. An axiomatic approach to risk measures was proposed by Artzner 
et al. \cite{Artzner1999Coherent} who introduced a set of postulates any coherent risk measure was required to satisfy. ES turned out to be coherent \cite{Acerbi2002Expected,Pflug2000Some} and was strongly advocated by academics. After a long debate, international regulation embraced it as the official risk measure in 2016 \cite{Basel2016Minimum}. 

The various risk measures discussed all involve averages. 
Since the distributions of financial data are not known the relative price movements of assets are observed at a number $T$ of time points and the true averages are replaced by empirical averages from these data. This works well if $T$ is sufficiently large, however, in addition to all the aforementioned problems, a general difficulty of portfolio optimization lies in the fact that the dimension $N$ of institutional portfolios (the number of different assets) is large, but the number $T$ of observed data per asset is never large enough, due to lack of stationarity of the time series and the natural limits (transaction costs, technical difficulties of rebalancing) on the sampling frequency. Therefore, portfolio optimization in large dimensions suffers from a high degree of estimation error, which renders the exercise more or less illusory, see e.g. \cite{Michaud1989TheMarkowitz}. Estimation of returns is even more error prone than the risk part, so several authors disregard the return completely, and seek the minimum risk portfolio, e.g. \cite{Kempf2006Estimating, Basak2009Jackknife, 
Frahm2010Dominating}. We follow the same approach here. 

In the two subsections that follow, we also assume that the returns are independent, symmetrically distributed random variables. This 
is, of course, not meant to be a realistic market model, but it allows us 
to make an explicit connection between the optimization of the portfolio variance under a constraint excluding short positions and the geometric problem of dichotomies discussed in section 2. This is all the more noteworthy because analytic results are notoriously scarce for portfolio optimization with no short positions.  We note that similar simplifying assumptions (Gaussian fluctuations, independence) were built into the original JP Morgan methodology, which was industry standard in its time, and influences the thinking of practitioners even today.

\subsection{Vanishing of the estimated variance}

We consider a portfolio of $N$ assets with weights $w_i,\;i=1,\dots, N$. The observations $r_i^t$ of the corresponding returns at various times $t=1,\dots, T$ are assumed to be independent, symmetrically distributed random variables. Correspondingly, the average value of the portfolio is zero. Its variance is given by 
\begin{equation}\label{defsi}
 \sigma_p^2=\frac{1}{T}\sum_t\left(\sum_i w_i r_i^t\right)^2
           =\sum_{i,j} w_i w_j\; \frac{1}{T}\sum_t r_i^t r_j^t=:\sum_{i,j} w_i w_j C_{ij},
\end{equation} 
where $C_{ij}$ denotes the covariance matrix of the observations. Note that the variance of a portfolio optimized in a given sample depends on the 
sample, so it is itself a random variable.

The variance of a portfolio obviously vanishes if the returns are fixed quantities that do not fluctuate. This subsection is not about such a trivial case. We shall see, however, that the variance optimized \emph{under a no-short constraint} can vanish with a certain probability if the dimension $N$ is larger than the number of observations $T$.

The rank of the covariance matrix is the smaller of $N$ and $T$ and for $N\le T$ the estimated variance is positive with probability one. Thus the optimization of variance can always be carried out as long as the number of observations $T$ is larger than the dimension $N$, albeit with a larger and larger error as $T/N$ decreases. For large $N$ and $T$ and fixed  $\alpha=T/N$ the estimation error increases as $\alpha/(\alpha-1)$ with decreasing $\alpha$ and diverges at $\alpha\downarrow 1$ \cite{Pafka2003Noisy, Burda2003Econophysics}. The divergence of the estimation error can be regarded as a phase transition. Below the critical value $\alpha_d:=1$ the optimization of variance becomes impossible. Of course, in practice one never has such an optimization task without some additional constraints. Note that because of the possibility of short selling (negative portfolio weights) the budget constraint (a hyperplane) in itself is not sufficient to forbid the appearence of large positive and negative positions, which then destabilize the optimization. In contrast, any constraint that makes the allowed weights finite can act as a regularizer.         
The usual regularizers are constraints on the norm of the portfolio vector. It was shown in \cite{Caccioli2013Optimal,Caccioli2016Lp}  how liquidity considerations lead in a natural way to regularization. Ridge regression (a constraint on the $\ell_2$ norm of the portfolio vector) prevents the convariance matrix from developing zero eigenvalues, and, especially in its nonlinear form \cite{Ledoit2012Nonlinear}, results in very satisfactory out-of-sample performance. 

An alternative is the $\ell_1$ regularizer, of which the exclusion of short positions is a special case. Together with the budget constraint, it prevents large sample fluctuations of the weights. Let us then impose the no-short ban, as it is indeed imposed in practice on a number of special portfolios (e.g. on pension funds), or, in episodes of crisis, on the whole 
industry. The ban on short selling extends the region where the variance can be optimized, but below $\alpha=1$ the optimization acquires a probabilistic character in that the regularized variance vanishes with a certain probability and the optimization can only be carried out when it is positive. (Otherwise there is a continuum of solutions, namely any combination of the eigenvectors belonging to zero eigenvalues, which makes the optimized variance zero.)

Interestingly, the probability of the variance vanishing is related to the problem of random dichotomies in the following way. For the portfolio variance \eqref{defsi} to become zero we need to have

\begin{equation}
 \sum_i w_i r_i^t=0
\end{equation} 
for all $t$. If we interchange $t$ and $i$ we see that according to \eqref{h3} this is possible as long as  the $N$ points in $\mathbb{R}^T$ with position vectors $\vec{r}_i:=\{r^t_i\}$ do not form a dichotomy. Hence the probability for zero variance is from \eqref{rescov} 

\begin{equation}\label{respzv}
 P_\mathrm{zv}(T,N)=1-P_\mathrm{d}(N,T)=1-\frac{1}{2^{N-1}}\sum_{i=0}^{T-1}\binom{N-1}{i}
   =\frac{1}{2^{N-1}}\sum_{i=T}^{N-1}\binom{N-1}{i}.
\end{equation} 
Therefore, the probability of the variance vanishing is almost 1 for small $\alpha$, decreases to the value 1/2 at $\alpha=1/2$, decreases further to 0 as $\alpha$ increases to 1 and remains identically zero for $\alpha>1$ \cite{Kondor2017Variancenoshort, Kondor2019Variancewithl1}. This is similar but also somewhat complementary to the curve shown in Fig.~\ref{fig:cover}. Formula \eqref{respzv} for the vanishing of the variance was first written up in \cite{Kondor2017Variancenoshort, Kondor2019Variancewithl1} on the basis of analogy with the minimax problem to be considered below, and it was also verified by extended numerical simulations. The above link to the Cover problem is a new result and it is rewarding to see how a geometric proof establishes a bridge between the two problems.

In \cite{Kondor2017Variancenoshort, Kondor2019Variancewithl1} an intriguing analogy with 
e.g. the condensed phase of an ideal Bose gas was pointed out. The analogous features are the vanishing of the chemical potential in 
the Bose gas, resp. the vanishing of the Lagrange multiplier enforcing the budget constraint in the portfolio problem; the onset of Bose condensation, resp. the appearence of zero weights ("condensation" of the solutions on the coordinate planes) due to the no-short constraint; the divergence of the transverse susceptibility, and the emergence of zero modes in both models.  
\subsection{The Maximal Loss}

The introduction of the Maximal Loss (ML) or minimax risk measure by Young \cite{Young1998AMinimax} in 1998 was motivated by numerical expediency. In contrast to the variance whose optimization demands a quadratic program, ML is constructed such that it can be optimized by linear programming, which could be performed very efficiently even on large datasets already at the end of the last century. Maximal Loss combines the worst outcomes of each asset and seeks the best combination of them. This may seem to be an over-pessimistic risk measure, but there are occasions when considering the worst outcomes is justifiable (think of an insurance portfolio in the time of climate change), and, as will be seen, the present regulatory market risk measure is not very far from ML. 

Omitting the portfolio's return again and focusing on the risk part, the maximal loss of a portfolio is given by 

\begin{equation}
 \text{ML}:=\min_{\vw}\max_{1\leq t\leq T} \left(-\sum_i w_i r_i^t\right)
\end{equation} 
with the constraint

\begin{equation}\label{simplconstr}
 \sum_i w_i=N.
\end{equation}
We are interested in the probability $P_\mathrm{ML}(T,N)$ that this minimax problem is feasible, that is ML does not diverge to $-\infty$. To this 
end we first eliminate the constraint \eqref{simplconstr} by putting 

\begin{equation}
 w_N=N-\sum_{i=1}^{N-1} w_i.
\end{equation} 
This results in 

\begin{equation}\label{ML2}
\text{ML}:=\min_{\tilde{\vw}}\max_{1\leq t\leq T}\left(-\sum_{i=1}^{N-1} w_i 
(r_i^t-r_N^t) -N r_N^t\right)
   =:\min_{\tilde{\vw}}\max_{1\leq t\leq T}\left(-\sum_{i=1}^{N-1} w_i \tilde{r}_i^{\,t} -N r_N^t\right)
\end{equation} 
with $\tilde{\vw}:=\{w_1,...,w_{N-1}\}\in\mathbb{R}^{N-1}$ and $\tilde{\vr}^{\,t}:=\{r^t_1-r^t_N,...,r^t_{N-1}-r^t_{N}\} \in\mathbb{R}^{N-1}$. 
For ML to stay finite for all choices of $\tilde{\vw}$ the $T$ random hyperplanes with normal vectors $\tilde{\vr}^t$ have to form a bounded cone. 
If the points $\tilde{\vr}^t$ form a dichotomy then according to \eqref{version1} there is a vector $\vW\in\mathbb{R}^{N-1}$ with $\vW\cdot\tilde{\vr}^t>0$ for all $t$. Since there is no constraint on the norm of $\tilde{\vw}$, the maximal loss \eqref{ML2} can become arbitrarily small for $\tilde{\vw}=\lambda\vW$ and $\lambda\to\infty$. The cone then is not bounded. We 
therefore find

\begin{equation}\label{resML}
 P_\mathrm{ML}(T,N)=P_\mathrm{d}(T,N-1)=\frac{1}{2^{T-1}}\sum_{i=0}^{N-2}\binom{T-1}{i}.
\end{equation} 
for the probability that ML cannot be optimized.

In the limit $N,T\to\infty$ with $\alpha=T/N$ kept finite \eqref{ML2} displays the same abrupt change as in the problem of dichotomies, a phase transition at $\alpha_c=2$. Note that this is larger than the critical point $\alpha_d=1$ of the unregularized variance, which is quite natural, since the ML uses only the extremal values in the data set. The probability for the feasibility of ML was first written up without proof in \cite{Kondor2007Noise} where a comparative study of the noise sensitivity of 
four risk measures, including ML, was performed. There are two important remarks we can make at this point. First, the geometric consideration above does not require any assumption about the data generating process; as long as the the returns are independent, they can be drawn from any symmetric distribution without changing the value of the critical point. This is a special case of the universality of critical points discovered by Donoho and Tanner \cite{Donoho2009Observed}.

The second remark is that the problem of bounded cones is closely related to that of bounded polytopes \cite{MinimaxProof}. The difference is just the additional dimension of the ML itself. If the random hyperplanes perpendicular to the vectors $\tilde{\vr}^t$ form a bounded cone for ML according to \eqref{ML2} then they will trace out a bounded polytope on hyperplanes perpendicular to the ML axis at sufficiently high values of ML. In fact, after the replacement $N-1\to N$ Eq.~\eqref{resML} coincides with the result in Theorem 4 of \cite{MinimaxProof} for the probability of $T$ random hyperplanes forming a bounded polytope in $N$ dimensions\footnote{There is a typo in Theorem 4 in \cite{MinimaxProof}; the summation has to start at $i=0$.}. The close relationship between the ML problem and the 
bounded polytope problem 
on the one hand, and the Cover problem on the other hand, was apparently not clarified before.

If we spell out the financial meaning of the above result, we are led to interesting ramifications. To gain an intuition, let us consider just two assets, $N=2$. If asset 1 produces a return sometimes above, sometimes below that of asset 2, then the minimax problem will have a finite solution. If, however, asset 1 dominates asset 2 (i.e. yields a return which is at least as large, and, at least at one time point, larger, than the return on asset 2 in a given sample), then, with unlimited short positions allowed, the investor will be induced to take an arbitrarily large long position in asset 1 and go correspondingly short in asset 2. This means that 
the solution of the minimax problem will run away to infinity, and the risk ML will be equal to minus infinity \cite{Kondor2007Noise}. The generalization to $N$ assets is immediate: if among the assets there is one that 
dominates the rest, or there is a combination of assets that dominates some of the rest, the solution will run away to infinity and ML will take the value of $-\infty$. This scenario corresponds to an arbitrage, the investor gains an arbitrarily large profit without risk \cite{Kondor2010Instability}. Of, course, if such a dominance is realized in one given sample, it may disappear in the next time interval, or the dominance relations can rearrange to display another mirage of an arbitrage.

Clearly, the ML risk measure is unstable against these fluctuations. In practice, such a brutal instability can never be observed, because there are always some constraints on the short positions, or groups of assets corresponding to branches of industry, geographic regions, etc. These constraints will prevent the instability to take place, the solution cannot run away to infinity, but will go as far as allowed by the constraints and 
then stick to the boundary of the allowed region. Note however that in such a case the solution will be determined more by the constraints (and ultimately by the risk manager imposing the constraints) rather than by the 
structure of the market. In addition, in the next period a different configuration can be realized, so the solution will jump around on the boundary defined by the constraints.

We may illustrate the role of short positions for the instability of ML further by investigating the case of portfolio weights $w_i$ that have to be larger than a threshold $\gamma\leq 0$. For $\gamma\to-\infty$ there are no restrictions on short positions whereas $\gamma=0$ correspondes to a complete ban on them. For $N,\,T\to \infty$ with $\alpha=T/N$ fixed 
the problem may be solved within the framework of statistical mechanics. The minimax problem for ML is equivalent to 
the following problem in linear programming: minimize the threshold variable $\kappa$ under the constraints \eqref{simplconstr}, $w_i\geq\gamma$, and 

\begin{equation}\label{linprogML}
 -\sum_i w_i r_i^t\leq \kappa \quad\forall t=1,...,T.
\end{equation} 
Similarly to \eqref{defGaEn} the central quantity of interest is

\begin{align}\label{defom}
	\Omega(\kappa, \gamma,\alpha)=\frac{ \int_{ \gamma}^{\infty} \prod_{i=1}^{N} dw_i \, \delta (\sum_{i}w_i-N ) \, \prod_{t=1}^{\alpha N} \Theta\left( \sum_{i} w_i r_i^t+\kappa \right)}{ \int_{ \gamma}^{\infty} \prod_{i=1}^{N} dw_i \, \delta ( \sum_{i}w_i-N )}
\end{align}
giving the fractional volume of points on the simplex defined by \eqref{simplconstr} that fulfill all constraints \eqref{linprogML}. For given $\alpha$ and $\gamma$ we decrease $\kappa$ down to the point $\kappa_c$ where the typical value of this fractional volume vanishes. The ML is then given by $\kappa_c(\alpha,\gamma)$. 

Some details of the corresponding calculations are given in the appendix. 
In Fig.~\ref{fig:resML} we show some results. As discussed above the divergence of ML for $\alpha<2$ is indeed formally eliminated for all $\gamma>-\infty$ and the functions $\text{ML}(\alpha;\gamma)$ smoothly interpolate between the cases $\gamma=0$ and $\gamma\to-\infty$. However, the situation is now even more dangerous since the unreliability of ML as risk measure for small $\alpha$ remains without being deducible from its divergence.

\begin{figure}[H]
	\centering
	\includegraphics[width=7.75cm]{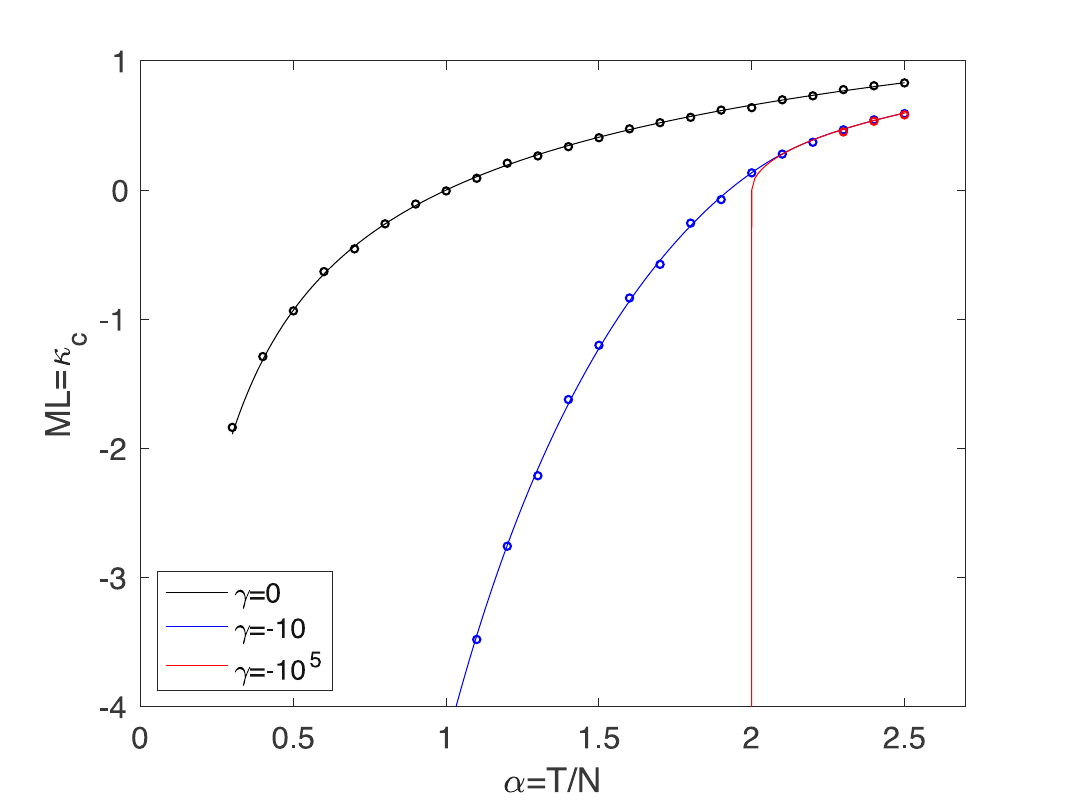}
	\includegraphics[width=7.75cm]{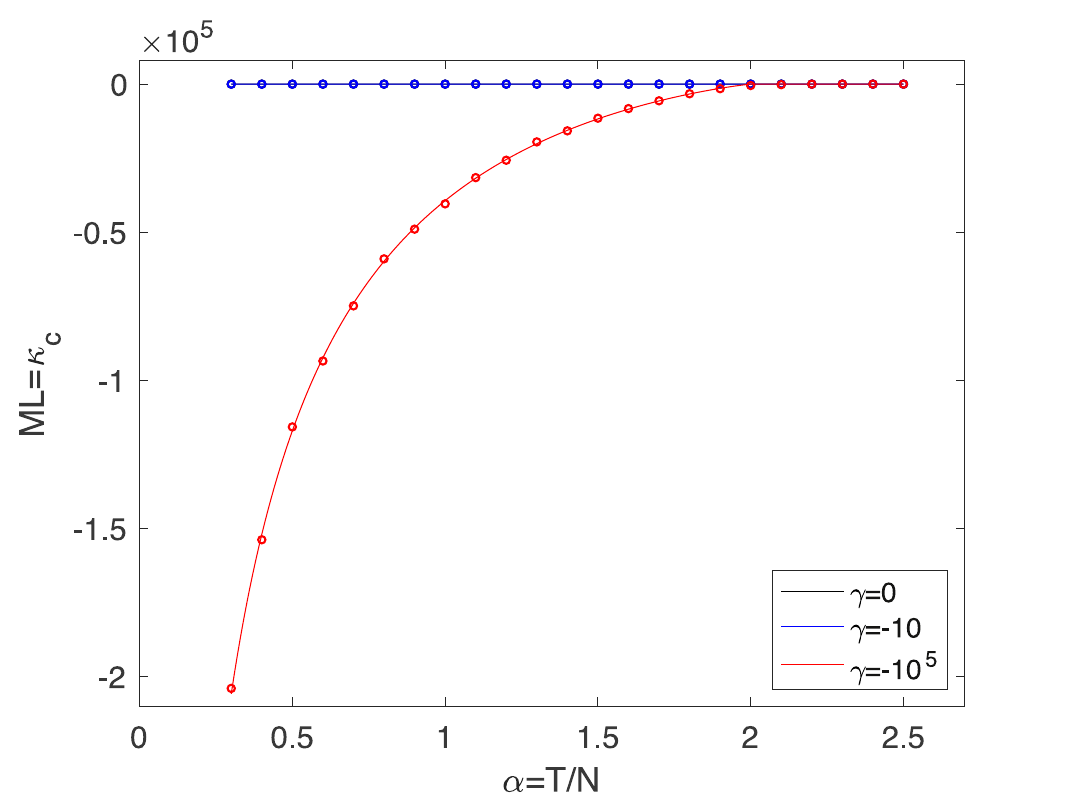}
	\caption{\textit{Left}: The Maximal Loss $\text{ML}=\kappa_c$ as a function of $\alpha$. The analytical results (solid line) are compared to simulation results (circles) with $N=200$ averaged over 100 samples. The symbol size corresponds to the statistical error. \textit{Right}: Same as 
left with largely extended axis of ML.\label{fig:resML}}
\end{figure}   

The recognition of the instability of ML as a dominance problem has proved very fruitful and led to a series of generalizations. First, it was realized \cite{Kondor2007Noise} that the instability of Expected Shortfall, of which ML is an extreme special case, has a very similar geometric origin. (The current regulatory ES is the expected loss above a 97.5\% quantile, whereas ML corresponds to 100\%.)  Both ES and ML are so called coherent risk measures \cite{Artzner1999Coherent}, and it was proved \cite{Kondor2010Instability} that the root of this instability lies in the coherence axioms themselves, so every coherent risk measure suffers from a similar instability. Furthermore, it was proved \cite{Kondor2010Instability} that the existence of a dominant-dominated pair of assets in the portfolio 
was a necessary and sufficient condition for the instability of ML, whereas it was only sufficient for other coherent risk measures. It follows that in terms of the variable $\alpha$ used in this paper (which is the reciprocal of the aspect ratio $N/T$ used in some earlier works, e.g. \cite{Ciliberti2007On, Kondor2010Instability, Vargahaszonits2008Instabilityofdownside}) the critical point of ML is a lower bound for the critical points of other coherent measures. Indeed, the critical line of ES was found to lie 
above the ML critical value of $\alpha_c=2$ \cite{Ciliberti2007On}. Value at Risk is not a coherent measure and can violate convexity, so it is not amenable to a similar study of its critical 
point. However, parametric VaR (that is the quantile where the underlying distribution is given, only its expectation value and variance is determined from empirical data) \emph{is} convex, and it was shown to possess a critical line that runs above that of ES \cite{Vargahaszonits2008Instabilityofdownside}. The investigation of the semi-variance yielded similar results \cite{Vargahaszonits2008Instabilityofdownside}. It seems then that the geometrical analysis of ML provides important information for a variety of risk measures, including some of the most widely used measures in the industry (VaR and ES) and also other downside risk measures. 

\section{Related problems}

In this section we list a few problems from different fields of mathematics and physics that are linked to the random coloring of points in high-dimensional space and point out their connection with the questions discussed above.

\subsection{Binary classifications with a perceptron}
Feed-forward networks of formal neurons perform binary classifications of input data \cite{HKP}. The simplest conceivable network of this type -- the perceptron -- consists of just an input layer of $N$ units $\xi_i$  and a single output bit $\zeta=\pm 1$ \cite{Rosenblatt}. Each input $\xi_i$ is directly connected to the output by a real valued coupling $w_i$. The output is computed as the sign of the weighted inputs

\begin{equation}\label{defperc}
 \zeta=\sign\left(\sum_{i=1}^N w_i\,\xi_i \right).
\end{equation} 
Consider now a family of random inputs $\{\xi_i^t\},\, t=1,...,T$ and ask for the probability $P_\mathrm{p}(T,N)$ that the perceptron is able to implement a randomly chosen binary classification $\{\zeta^t\}$ of these inputs. Interpreting the vectors $\vxi^t:=\{\xi_i^t\}$ as position vectors of $T$ points in $N$ dimensions and the required classifications $\zeta^t$ as a black-white coloring we hence need to know the probability that this particular coloring is a dichotomy. Indeed, if a hyperplane exists 
that separates black points from white ones it has a normal vector $\vw$ that gives a suitable choice for the perceptron weights to get all classifications right. Therefore we have 

\begin{equation}\label{resperc}
 P_\mathrm{p}(T,N)=P_\mathrm{d}(T,N)=\frac{1}{2^{T-1}}\sum_{i=0}^{N-1}\binom{T-1}{i}.
\end{equation}
In the thermodynamic limit $N, T\to \infty$ this problem together with a variety of modifications can be analyzed using methods from the statistical mechanics of disordered systems along the lines of Eqs.~\eqref{defGaEn}-\eqref{resGaEn}, see \cite{EnvB}.

\subsection{Zero-sum games with random pay-off matrices}
In game theory two or more players choose among different strategies at their disposal and receive a pay-off (that may be negative) depending on the choices of all participating players. A particularly simple situation is given by a zero-sum game between two players where one players profit is the other players loss. If the first player may choose among $N$ strategies and the second among $T$ the setup is defined by an $N\times T$ pay-off matrix $r_i^t$ giving the reward for the first player if he plays strategy $i$ and his opponent strategy $t$. Barring rare situations in which it is advantageous for one or both players to always choose one and the same strategy, it is known from the classical work of Morgenstern and von 
Neumann \cite{MovN} that the best the players can do is to choose at random with different probabilities among their available strategies. The set of these probabilities $p_i$ and $q^t$, respectively, is called a mixed strategy.

For large numbers of available strategies it is sensible to investigate typical properties of such mixed strategies for random pay-off matrices. This can be done in a rather similar way to the calcuation of ML presented in the appendix of the present paper \cite{BeEn}. One interesting result is, that an extensive part of the probabilities $p_i$ and $q^t$ forming the optimal respective mixed strategies have to be identically zero: for both players there are strategies they should never touch.

\subsection{Non-negative solutions to large systems of linear equations}

Consider a random $N\times T$ matrix $r_i^t$  and a random vector $\vb\in \mathbb{R}^N$. When will the system of linear equations

\begin{equation}\label{linsys}
 \sum_t r_i^t x^t=b_i, \qquad i=1,...,N
\end{equation}
typically have a solution with all $x^t$ being non-negative? This question is 
related to the optimization of financial portfolios under a ban of short selling as discussed above and also occurs when investigating the stability of chemical or ecological problems \cite{MacArthur, May}. Here the $x^t$ denote concentrations of chemical or biological species and hence have to be non-negative. Similar to optimal mixed strategies considered in the previous subsection the solution has typically a number of entries $x^t$ that are strictly zero (species that died out) and the remaining ones being positive (surviving species). And again for $T=\alpha N$ and $N\to\infty$ a sharp transition at a critical value $\alpha_c$ that separates situations with typically no non-negative solution from those in which typically such a solution can be found \cite{LaEn}. 

To make contact with the cases discussed before it is useful to map the problem to a dual one by using again Farkas' lemma. Let us denote by 

\begin{equation}
 \bar\vr=\sum_t c^t \vr^t, \qquad c^t\geq 0 ,\;t=1,...,T
\end{equation} 
the vectors in the non-negative cone of the column vectors $\vr^t$ of matrix $r_i^t$.  It is clear that \eqref{linsys} has a non-negative solution 
$\vx$ if $\vb$ belongs to this cone and that no such solution exists if $\vb$ lies outside the cone. In the latter case, however, there must be a hyperplane separating $\vb$ from the cone. Denoting the normal of this hyperplane by $\vw$ we have hence the following duality: Either the system \eqref{linsys} has a non-negative solution $\vx$ or there exists a vector 
$\vw$ with

\begin{equation}\label{dual}
 \vw\cdot\vr^t\geq 0 \quad t=1,...,T \qquad\mathrm{and}\qquad \vw\cdot\vb<0.
\end{equation} 
If the $r^t_i$ are drawn independently from a distribution with finite first and second cumulant $R$ and  $\sigma_r^2$, respectively, and the components $b_i$ are independent random numbers with average $B$ and variance 

$\sigma_b^2/N$ the dual problem \eqref{dual} may be analyzed along the lines of \eqref{defGaEn}-\eqref{resGaEn}. The result for the typical entropy of solution vectors $\vw$ reads \cite{LaEn}

\begin{equation}\label{resdualEn}
 S_\mathrm{typ}(\gamma,\alpha)
     =\extr_{q,\kappa}\left[\frac{1}{2}\ln(1-q)+\frac{q}{2(1-q)}-\frac{\kappa^2\gamma}{2(1-q)}
      +\alpha\int Dt \ln H\left(\frac{\kappa-\sqrt{q}t}{\sqrt{1-q}}\right)\right],
\end{equation} 
where the parameter

\begin{equation}
 \gamma:=\left(\frac{B\,\sigma_r}{R\,\sigma_b}\right)^2
\end{equation} 
characterizes the distributions of $r^t_i$ and $b_i$. The main difference to \eqref{resGaEn} is the additional extremum over $\kappa$ regularized by the penalty term proportional to $\kappa^2$. Considering the limit $q\to 1$ in \eqref{resdualEn} it is possible to determine the critical value $\alpha_c(\gamma)$ bounding the region where typically no solution $\vw$ may be found. For nonrandom $\vb$, i.e. $\sigma_b\to0$ implying $\gamma\to\infty$ we find back the Cover result $\alpha_c=2$.

The problem is closely related to a phase transition found recently in MacArthur's resource competition model \cite{MacArthur,TiMo,LaEn} in which a community of purely competing species builds up a collective cooperative phase above a critical threshold of the biodiversity.


\section{Discussion}

We have reviewed in this paper various problems from different disciplines, including high dimensional random geometry, finance, binary classification with a perceptron, game theory, and random linear algebra, which all 
have at their root the problem of dichotomies, that is the linear separability of points carrying a binary label and scattered randomly over a high-dimensional space. No doubt there are several further problems belonging to this class; those that spring to mind are theoretical ecology alluded to at the end of the previous Section, or linear programming with random parameters \cite{MinimaxTodd}. Some of these conceptual links are obvious and have been known for decades (for example the link between dichotomies and the perceptron), others are far less clear at first sight, such as e.g. the relationship with the two finance problems discussed in Sec. 3. We regard as one of the merits of this paper the establishment of this network of conceptual connections between seemingly faraway areas of study. Apart from the occasional use of the heavy machinery of replica theory, in most of the paper we offered transparent geometric arguments, where basically our only tool was the Farkas' lemma.

The phase transitions we encountered in all of the problems discussed here are similar in spirit to the geometric transitions discovered by Donoho 
and Tanner \cite{Donoho2009Observed} and interpreted at a very high level 
of abstraction by \cite{Amelunxen2013Living}. One of the central features 
of these transitions is the universality of the critical point. This universality is different from the one observed in the vicinity of continuous 
phase transitions in physics, where the value of the critical point can widely vary even between transitions belonging to the same universality class. The universality in physical phase transitions is a property of the critical indices and other critical parameters. Critical indices appear also in our abstract geometric problems, and they are universal, but we omitted their discussion which might have led far from the main theme.

At the bottom of our geometric problems there is the optimization of a convex objective function (which is, by the way, the key to the replica symmetric solutions we found). The recent evolution of neural networks, machine learning and artificial intelligence is mainly concerned with a radical lack of convexity, which points to the direction in which we may try to extend our studies. Another simplifying feature we exploited was the independence of the random variables. The moment correlations appear, these problems become hugely more complicated. We left this direction for future exploration. However, it is evident that progress in any of these problems will induce progress in the other fields, and we feel that  having revealed their fundamental unity may help the transfer of methods and ideas between these fields. This may be the most important achievement of this analysis.

\vspace{6pt}

\authorcontributions{Conceptualization, I.K. and A.E.; formal analysis, A.P., I. K. and A. E.; software, A. P.; writing – original draft, A.P., I. K. and A. E..}


\acknowledgments{A.P. and A.E. are grateful to Stefan Landmann for many interesting discussions.}


\appendixtitles{yes} 


\appendix
\section{Replica calculation of maximal loss}
In this appendix, we provide some details for the determination of the maximal loss of a random portfolio using the replica trick. The calculation is a generalization of the one presented in \cite{BeEn} for random zero-sum games. A presentation at full length can be found in \cite{axel}. As we pointed out in the main text, maximal loss is a special limit of the Expected Shortfall risk measure, corresponding to the so called confidence level going to 100\%. In \cite{Caccioli2018Portfolio} a detailed study of the behavior of  ES was carried out, including the limiting case of maximal loss. That treatment is completely different from the one in here, so the present calculation can be regarded as complementary to that in \cite{Caccioli2018Portfolio}.

The central quantity of interest is the fractional volume

\begin{align}
	\Omega(\kappa, \gamma,\alpha)=\frac{ \int_{ \gamma}^{\infty} \prod_{i=1}^{N} dw_i \, \delta (\sum_{i}w_i-N ) \, \prod_{t=1}^{\alpha N} \Theta\left( \sum_{i} w_i r_i^t+\kappa \right)}{ \int_{ \gamma}^{\infty} \prod_{i=1}^{N} dw_i \, \delta ( \sum_{i}w_i-N )}
\end{align}
defined in \eqref{defom}. Although not explicitly indicated, $\Omega(\kappa,\gamma,\alpha)$ depends on all the random parameters $r_i^t$ and is therefore by itself a random quantity. The calculation of its complete probability density $P(\Omega)$ is hopeless but for large $N$ this distribution gets concentrated around the typical value $\Omega_\mathrm{typ}(\kappa,\gamma,\alpha)$. Because $\Omega$ involves a product of many independent 
random factors this typical value is given by 

\begin{align}
	\Omega_\mathrm{typ}(\kappa, \gamma,\alpha)
        =e^{\langle\langle\ln \Omega(\kappa, \gamma,\alpha)\rangle\rangle}
\end{align} 
rather than by $\langle\langle\Omega(\kappa,\gamma,\alpha) \rangle\rangle$. Here $\langle\langle ...\rangle\rangle$ denotes the average over the $r_i^t$. A direct calculation of $\langle\langle \ln\Omega\rangle\rangle$ is hardly possible. It may be circumvented by exploiting the identity

\begin{align}
	\langle\langle \ln (\Omega(\kappa, \gamma,\alpha)) \rangle\rangle=\lim\limits_{n \rightarrow 0} \frac{1}{n} \left[ \langle\langle \Omega^n(\kappa, \gamma,\alpha) \rangle\rangle -1 \right]
\end{align}
For natural $n$ the determination of $\langle\langle \Omega^n\rangle\rangle$ is feasible. The main problem then is to continue the result to real $n$ in order to perform the limit $n\to 0$. 

The explicit calculation starts with

\begin{align}
	\langle\langle \Omega(\kappa, \gamma, \alpha)^n   \rangle\rangle= \bigg\langle\bigg\langle \frac{  \int_{ \gamma}^{\infty} \prod_{i=1}^{N} \prod_{a=1}^{n} dw_i^a \, \prod_{a=1}^n \delta (\sum_{i}w_i^a-N  ) \, \prod_{t=1}^{\alpha N} \prod_{a=1}^{n} \Theta( \sum_{i} w_i^a r_i^t+\kappa )}{\int_{ \gamma}^{\infty} \prod_{i=1}^{N} \prod_{a=1}^{n} dw_i^a \, \prod_{a=1}^n \delta ( \sum_{i}w_i^a-N )} \bigg\rangle\bigg\rangle.
\end{align}
Using 

\begin{align}
 \int_{ \gamma}^{\infty} \prod_{i=1}^{N} dw_i \, \delta ( \sum_{i}w_i-N ) \sim \exp\{N[1+\ln(1-\gamma)]\}
\end{align}
for large $N$ and representing the $\delta$-functions and $\Theta$-functions by integrals over auxiliary variables $E_a, \lambda_t^a$, and $y_t^a$ we arrive at

\begin{align}
	\begin{split}
	\langle\langle \Omega(\kappa, \gamma, \alpha)^n   \rangle\rangle=& \exp\left\lbrace  -nN\left[ 1+ \ln (1- \gamma)\right]\right\rbrace \\
	& \times \int_{\gamma}^{\infty} \prod_{i,a} dw_i^a \, \int \prod_{a} \frac{dE_a}{2 \pi}  \, \exp\left[ iN \sum_{a} E_a\left( \frac{1}{N}\sum_{i}w_i^a-1\right)  \right] \\\label{h1}
	& \times \int_{-\kappa}^{\infty} \prod_{t,a} d\lambda_t^a \int \prod_{t,a} \frac{dy_t^a}{2 \pi}  \exp  \left( i \sum_{t,a} y_t^a \lambda_t^a\right)  \bigg\langle\bigg\langle  \exp  (- i\sum_{i,t,a} y_t^a w_i^a r_i^t)\bigg\rangle\bigg\rangle.
\end{split}
\end{align}
The average over the $r_i^t$ may now be performed for independent Gaussian $r_i^t$ with average zero and variance $\sigma^2=1/N$. The result is valid also for more general distributions. First, multiplying the variance by a constant just rescales the maximal loss but does not influence the 
optimal $\vw$. Second, 
for $N\to\infty$ only the first two cumulants of the distribution matter due to the central limit theorem. 
Crucial is, however, the assumption of the $r_i^t$ being independent.

Performing the average we find

\begin{align}\nonumber
	\biggl\langle\biggl\langle \exp  \left( - i\sum_{i,t,a}y_t^a w_i^a r_i^t 
\right)  \biggr\rangle \biggr\rangle=&\prod_{i,t} \left[  \int \frac{dr_i^t}{\sqrt{2 \pi \sigma^2}}  \exp  \left( -\frac{(r_i^t)^2}{2\sigma^2} 
- i r_i^t \sum_{a} y_t^a w_i^a  \right) \right] \\
	=&\exp  \left( -\frac{1}{2N} \sum_{i,t} \sum_{a,b} w_i^a w_i^b y_t^a y_t^b\right).
\end{align}
To disentangle in \eqref{h1} the $w$-integrals from those over $\lambda$ and $y$ we introduce the order parameters 

\begin{align}
	q_{ab}=\frac{1}{N}\sum_{i}w_i^aw_i^b,  \quad a\geq b
\end{align} 
together with the conjugate ones $\hat{q}_{ab}$. Using standard techniques \cite{EnvB} we end up with

\begin{align}
	\begin{split}
	\langle\langle \Omega(\kappa, \gamma, \alpha)^n   \rangle\rangle=&    \int \prod_{a \geq b} \frac{dq_{ab} d\hat{q}_{ab}}{2 \pi/N} \, \int \prod_{a} \frac{dE_a}{2 \pi} \\\label{h2}
	&\times \exp \left\lbrace -iN \sum_{a\geq b}q_{ab} \hat{q}_{ab} -iN \sum_{a}E_a - nN  \left[ 1+ \ln\left( 1- \gamma\right) \right] + N G_s +\alpha NG_E\right\rbrace,
\end{split}
\end{align}
where 

\begin{align}\label{defGS}
	G_S=&\ln\left[ \int_{ \gamma}^{\infty} \prod_{a}  dw^a \exp\left(i \sum_{a\geq b} \hat{q}_{ab} w^aw^b + i \sum_{a} E_a w^a \right)\right]
\end{align}
and

\begin{align}\label{defGE}
	G_E=&\ln \left[ \int_{-\kappa}^{\infty} \prod_{a}  d\lambda^a \int \prod_{a}   \frac{dy^a}{2 \pi} \exp  \left(-\frac{1}{2}\sum_{a,b}q_{ab} y^ay^b + i \sum_{a} y^a \lambda^a\right)\right].
\end{align}
For $N\to\infty$ the integrals over the order parameters in \eqref{h2} may be calculated using the saddle-point method. The essence of the so-called replica-symmetric ansatz is the assumption that the values of the order parameters at the saddle-point are invariant under permutation of the replica indices $a$ and $b$. In \cite{axel} arguments are given why the replica-symmetric saddle-point should yield correct results in the present context. We therefore assume for the saddle-point values of the order parameters

\begin{equation}
 \begin{split}
  q_{aa}&= \, q_1 \quad \quad i\hat{q}_{aa}=-\frac{1}{2}\hat{q}_1 \quad \quad 
    iE_a=E \quad \quad \forall a\\
  q_{ab}&= \, q_0 \quad \quad i\hat{q}_{ab}=\hat{q}_0 \hspace{3.1cm} \forall a>b.
 \end{split}
\end{equation}
which implies various simplifications in \eqref{h2}-\eqref{defGE}. Employing standard manipulations \cite{EnvB} we arrive at

\begin{align}
	\langle\langle \Omega(\kappa, \gamma, \alpha)^n \rangle\rangle \sim \exp \left\lbrace N \underset{q_0,\hat{q}_0,q_1,\hat{q}_1,E}{\text{extr}} \, \left[ -\frac{n(n-1)}{2}q_0\hat{q}_0+\frac{n}{2}q_1\hat{q}_1 -nE - n  (1+ \ln(1- \gamma))  +  G_S +\alpha G_E\right] \right\rbrace.
\end{align}
Using the shorthand notations \eqref{shorthand} the functions $G_S$ and $G_E$ are now given by

\begin{align}
	G_S=&\ln \int Dl \left[  \exp\left( \frac{(\sqrt{\hat{q}_0}l+E)^2}{2(\hat{q}_0+\hat{q}_1)}\right) \sqrt{\frac{2 \pi}{\hat{q}_0+\hat{q}_1}} H\left(- \frac{\sqrt{\hat{q}_0}l+E-\gamma (\hat{q}_0+\hat{q}_1)}{\sqrt{\hat{q}_0+\hat{q}_1}}\right) \right] ^n
\end{align}
and

\begin{align}
	G_E=\ln  \int Dm \, H\left( \frac{\sqrt{q_0} m - \kappa}{\sqrt{q_1 - q_0}}\right)^n .
\end{align}
We may now treat $n$ as a real number and perform the limit $n\to 0$. In this way we find for the averaged entropy

\begin{align}
	S(\kappa, \gamma,\alpha):=& \lim_{N \rightarrow \infty} \frac{1}{N} \langle\langle \ln \left[ \Omega(\kappa, \gamma,\alpha)\right]   \rangle\rangle
	=\lim_{N \rightarrow \infty} \frac{1}{N}  \lim\limits_{n \rightarrow 0} \frac{1}{n} \left[ \langle\langle \Omega(\kappa, \gamma,\alpha)^n  \rangle\rangle -1 \right]
\end{align}	
the expression

\begin{align}
	\begin{split}
	S(\kappa, \gamma, \alpha)= \underset{q_0,\hat{q}_0,q_1,\hat{q}_1,E}{\text{extr}}&\left[   \frac{q_0\hat{q}_0}{2}+\frac{q_1\hat{q}_1}{2} -E - 1- \ln(1- \gamma)  +\frac{1}{2}\ln(2\pi) - \frac{1}{2}\ln(\hat{q}_0+\hat{q}_1)  \right.\\
	&\left. +\frac{ \hat{q}_0 + E^2}{2(\hat{q}_0+\hat{q}_1)}
	+ \int Dl \ln H\left(- \frac{\sqrt{\hat{q}_0}l+E-\gamma (\hat{q}_0+\hat{q}_1)}{\sqrt{\hat{q}_0+\hat{q}_1}}\right) \right.\\
	&\left.   +\alpha \int Dm \ln H\left( \frac{\sqrt{q_0}m-\kappa}{\sqrt{q_1-q_0}}\right) \right].
	\label{entropie} 
\end{split}  
\end{align}
The remaining extremization has to be done numerically. Before embarking on this task it is useful to remember that $\Omega$ and $S$ are only instrumental in determining the maximal loss which in turn is given by the value $\kappa_c$ of $\kappa$ for which $\Omega$ tends to zero. At the same time the typical overlap $q_0$ between two different vectors in $\Omega$ has to tend to the self-overlap $q_1$. To investigate this limit we replace the order parameter $q_1$ by

\begin{equation}\label{defv}
 v:=q_1-q_0
\end{equation}
and study the saddle-point equations for $v\to 0$. In this limit it turns out that the remaining order parameters may either also tend to zero or diverge. It is therefore convenient to make the replacements

\begin{equation}\label{defresop}
 \hat{q}_0\to\frac{\hat{q}_0}{v^2},\quad \hat{q}_1\to \hat{w}:=\frac{\hat{q}_1+\hat{q}_0}{v},\quad
  E\to\frac{E}{v}.
\end{equation} 
Rescaled in this way the saddle-point values of the order parameters remain $\mathcal{O}(1)$ for $v\to 0$. After some tedious calculations the saddle-point equations acquire the form

\begin{equation}\label{saddle}
 \begin{split}
  &0=\hat{w}- \alpha H\left( \frac{\kappa_c}{\sqrt{q_0} }\right) \\
  &0=-\hat{q_0}+\hat{w}(q_0+\kappa_c^2)-\alpha \sqrt{q_0} \kappa_c\,  
      G\left( \frac{\kappa_c}{\sqrt{q_0} }\right)\\
  &0=E(1-\gamma)-\hat{w}(q_0-\gamma)+\hat{q}_0\\ 
  &0=\hat{w}-H\left(-\frac{E - \gamma \hat{w}}{\sqrt{\hat{q}_0}}\right)\\
  &0=\hat{w}(E-1)+\sqrt{\hat{q}_0}\,G\left(\frac{E - \gamma \hat{w}}{\sqrt{\hat{q}_0}}\right)
      +\gamma \hat{w}(1-\hat{w})
 \end{split}
\end{equation}
where

\begin{align}
G(x):=&\frac{1}{\sqrt{2 \pi}} e^{-\frac{x^2}{2}}.
\end{align}
From the numerical solution of the system \eqref{saddle} we determine $\kappa_c(\alpha,\gamma)$ as shown in Fig.~\ref{fig:resML}.

\reftitle{References}


\externalbibliography{yes}
\bibliography{paper.bib}


\end{document}